ORDER, DISORDER, AND PHASE TRANSITION
IN CONDENSED SYSTEM

# BCS–BEC Crossover and Quantum Hydrodynamics in *p*-Wave Superfluids with a Symmetry of the *A*1 Phase[1]

**M. Yu. Kagan[a] and D. V. Efremov[a,b,c]**

[a] *Kapitza Institute for Physical Problems, Moscow, 119334 Russia*
[b] *Max-Planck-Institute for Solid State Research, Stuttgart, D-70569 Germany*
[c] *Max-Planck-Institut für Physik Komplexer Systeme, Dresden, 01187 Germany*
e-mail: kagan@kapitza.ras.ru
Received July 14, 2009

**Abstract**—We solve the Leggett equations for the BCS–BEC crossover in a three dimensional resonance *p*-wave superfluid with the symmetry of the *A*1 phase. We calculate the sound velocity, the normal density, and the specific heat for the BCS domain ($\mu > 0$), for the BEC domain ($\mu < 0$), and close to the important point $\mu = 0$ in the 100% polarized case. We find the indications of a quantum phase transition close to the point $\mu(T = 0) = 0$. Deep in the BCS and BEC domains, the crossover ideas of Leggett, Nozieres, and Schmitt–Rink work quite well. We discuss the spectrum of orbital waves, the paradox of intrinsic angular momentum and the complicated problem of chiral anomaly in the BCS *A*1 phase at $T = 0$. We present two different approaches to the chiral anomaly, based on supersymmetric hydrodynamics and on the formal analogy with the Dirac equation in quantum electrodynamics. We evaluate the damping of nodal fermions due to different decay processes in the superclean case at $T = 0$ and find that a ballistic regime $\omega\tau \gg 1$ occurs. We propose to use aerogel or nonmagnetic impurities to reach the hydrodynamic regime $\omega\tau \ll 1$ at $T = 0$. We discuss the concept of the spectral flow and exact cancelations between time derivatives of anomalous and quasiparticle currents in the equation for the total linear momentum conservation. We propose to derive and solve the kinetic equation for the nodal quasiparticles in both the hydrodynamic and ballistic regimes to demonstrate this cancelation explicitly. We briefly discuss the role of the other residual interactions different from damping and invite experimentalists to measure the spectrum and damping of orbital waves in the *A* phase of $^3$He at low temperatures.

**DOI:** 10.1134/S1063776110030064

## 1. INTRODUCTION

The first experimental results on the *p*-wave Feshbach resonance [1–3] in ultracold fermionic gases $^{40}$K and $^6$Li make the field of quantum gases closer to the interesting physics of superfluid $^3$He and the physics of unconventional superconductors such as $Sr_2RuO_4$. In this context, it is important to bridge the physics of ultracold gases and the low-temperature physics of quantum liquids and anomalous superconductors and thus to enrich both communities with the experience and knowledge accumulated in each of these fields. The purpose of this paper is first and foremost to describe the transition from the weakly bound Cooper pairs with a *p*-wave symmetry to strongly bound local *p*-wave pairs (molecules) and to try to reveal the nontrivial topological effects related to the presence of nodes in the superfluid gap of the 100%-polarized *p*-wave *A*1 phase in three dimensions. We note that the *A*1 phase symmetry is relevant both to ultracold Fermi gases in the *p*-wave Feshbach resonance regime and to super-fluid $^3$He-*A* in the presence of a large magnetic field or a large spin polarization. We give a special attention to the spectrum of collective excitations and to the superfluid hydrodynamics of the *A*1 phase at $T = 0$, where topological effects are very pronounced, especially in the BCS domain. We propose an experimental verification of the different approaches related to the complicated problem of chiral anomaly and the mass-current nonconservation in the superfluid *A*1 phase of $^3$He in the superclean case and in the presence of aerogel as well as for the *A*1 *p*-wave condensates in magnetic traps in the presence of Josephson tunneling currents.

This paper is organized as follows. Section 1 provides an introduction. In Section 2, we briefly comment on recent experiments on the *p*-wave Feshbach resonance and describe the global phase diagram for 100%-polarized *p*-wave resonance superfluids in 3D. In Section 3, we describe the quasiparticle spectrum and nodal points in the *A*1 phase. In Section 4, we solve mean-field Leggett equations for triplet superfluids with the symmetry of the *A*1 phase at $T = 0$ and study the behavior of the superfluid gap $\Delta$, the chemical potential $\mu$, and the sound velocity $c_S$ deep in the BCS ($\mu > 0$) and BEC domains ($\mu < 0$) as well as close to the interesting point $\mu = 0$. In Section 5, we study the temperature behavior of the normal density $\rho_n$ and specific heat $C_v$ in the BCS domain, in the BEC domain, and close to $\mu = 0$, where we find indications of a quantum phase transition. In Section 6, we describe the orbital wave spectrum in the BCS and BEC domains of the $A$1 phase and describe the complicated problem of chiral anomaly (mass-current nonconservation) in the superfluid hydrodynamics of the $A$1 phase in the BCS domain at $T \longrightarrow 0$. In Sections 7 and 8, we present two different approaches to the calculation of the anomalous current: based on supersymmetric hydrodynamics [4] and on the analogy with the Dirac equation in quantum electrodynamics (QED) [5, 6]. We note that both approaches are very general. The first is based on the inclusion of the fermionic Goldstone mode in the low-frequency hydrodynamic action [4]. It can be useful for all nodal superfluids and superconductors with zeroes of the superconductive gap, such as $^3$He-$A$, $Sr_2RuO_4$, $UPt_3$, $UNi_2Al_3$, and $U_{1-x}Th_xBe_{13}$ [7]. The second approach is also very nice and general. It is connected with the appearance of the Dirac-like spectrum of fermions with a zero mode [5, 6], which also arises in many condensed-matter systems such as $^3$He-$A$, chiral superconductor $Sr_2RuO_4$, organic conductor $\alpha$-(BEDT-TTF)$_2$I$_3$, 2D semiconductors, or recently discovered graphene [7–10]. In Section 9, we evaluate the damping in the superclean $A$1 phase at $T = 0$ due to different decay processed and conclude that the ballistic regime $\omega\tau \gg 1$ occurs at $T = 0$. We propose to use aerogel or nonmagnetic impurities to reach the hydrodynamic regime $\omega\tau \ll 1$. We discuss the concept of the spectral flow and exact cancelations of anomalies between time derivatives of the anomalous and quasiparticle currents in the equation of the total linear momentum conservation. We also propose to derive a kinetic equation for nodal quasiparticles in both hydrodynamic and ballistic regimes and to demonstrate this cancelation explicitly. In Section 10, we provide our conclusions and acknowledgments. We also invite experimentalists to measure the spectrum and damping of the orbital waves in the $^3$He-$A$ phase at low temperatures $T \ll T_C$ and thus to help resolve the orbital momentum paradox. We also propose to extend the measurements of the orbital inertia and the orbital viscosity in nonsingular vortex textures in the conservation $A$ phase [11] to low temperatures via creating spin polarization. Finally, we propose to measure the Josephson current between two 2D films of the axial and planar phases with an attempt to directly extract the difference between topological charges $\Delta Q = 1$ in these phases.

---

[1] The article is published in the original.

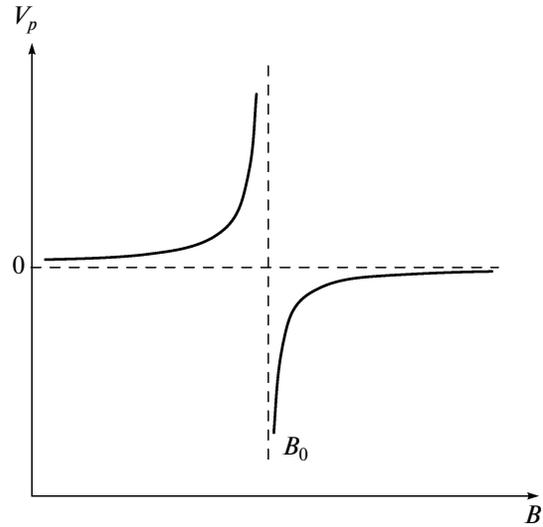

**Fig. 1.** Sketch of the $p$-wave Feshbach resonance. The scattering volume $V_p$ diverges at $B = B_0$.

## 2. FESHBACH RESONANCE AND PHASE DIAGRAM FOR 100%-POLARIZED $p$-WAVE RESONANCE SUPERFLUIDS

In the first experiments on the $p$-wave Feshbach resonance, experimentalists measured the molecule formation in the ultracold fermionic gas of $^6$Li atoms close to the resonance magnetic field $B_0$ [1, 2].

In the last years, analogous experiments on the $p$-wave molecule formation in the spin-polarized fermionic gas of $^{40}$K-atoms were started [3]. The lifetime of $p$-wave molecules is still rather short [1–3]. But the physicists working in ultracold gases have started intensively studying the huge bulk of experimental and theoretical wisdom accumulated in the physics of superfluid $^3$He and anomalous complex superconductors (see [12, 13]).

To understand the essence of the $p$-wave Feshbach resonance, we recall the basic formula for the $p$-wave scattering amplitude in the vacuum (see [14, 15])

$$f_{l=1}(E) = \frac{pp'}{\dfrac{1}{V_p} + \dfrac{2mE}{\pi r_0} + i(2mE)^{3/2}}, \quad (1)$$

where $l = 1$ is the orbital momentum in the $p$-wave channel, $E$ is the two-particle energy, $V_p = r_0^2 a_p$ is the scattering volume, $a_p$ is the $p$-wave scattering length, $r_0$ is the interaction range, and $p$ and $p'$ are the incoming and outgoing momenta. For the Feshbach resonance in fermionic systems, $p \sim p' \sim p_F$ and usually $p_F r_0 < 1$. The scattering length $a_p$ and hence the scattering volume $V_p$ diverges in the resonance magnetic field $B_0$ (see Fig. 1), $1/V_p = 1/a_p = 0$. The imaginary part of the scattering amplitude $f_p$ is small and nonzero only for positive energies $E > 0$, and hence the $p$-wave Feshbach resonance is intrinsically narrow. We note that for

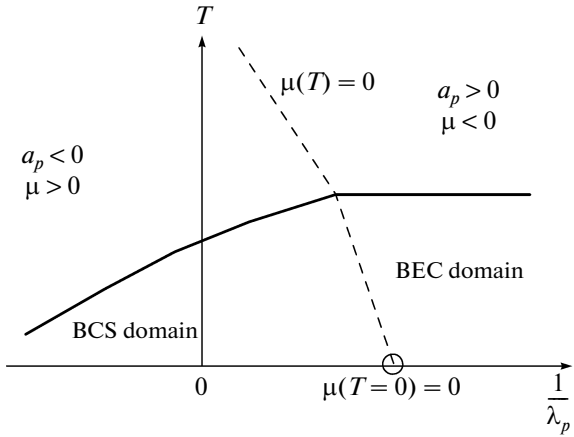

**Fig. 2.** Qualitative picture of the BCS–BEC crossover in the 100%-polarized $A1$ phase for $p$-wave superfluids. We indicate the line where $\mu(T) = 0$ and the quantum phase-transition point $\mu(T=0) = 0$.

negative energies $E < 0$, there is a molecular bound state:

$$|E_b| = \frac{\pi r_0}{2mV_p} = \frac{\pi}{2mr_0 a_p}.$$

In the unitary limit, the molecular binding energy $|E_b| \longrightarrow 0$.

The first theoretical articles on the $p$-wave Feshbach resonance often deal with a mean-field two-channel description of the resonance [15]. In this paper, we study the $p$-wave Feshbach resonance in the framework of a one-channel description, which is closer to the physics of superfluid $^3$He and captures the essential physics of the BCS–BEC crossover in $p$-wave superfluids rather well.

In magnetic traps (in the absence of the so-called dipolar splitting), the fully (100%) polarized gas or, more precisely, one hyperfine component of the gas is usually studied. In the language of $^3$He, the pairs with $S_{\text{tot}} = S_z^{\text{tot}} = 1$, or $|\uparrow\uparrow\rangle$-pairs are studied. In this paper, we consider the $p$-wave triplet $A1$ phase in three dimensions with $S_{\text{tot}} = S_z^{\text{tot}} = 1$.

A qualitative picture of the global phase diagram of the BCS–BEC crossover in the 100%-polarized $A1$ phase is presented in Fig. 2. In its gross features, it resembles the phase diagram of the BCS–BEC crossover for $s$-wave pairing (see [16] for more details). However, there is a very interesting question about the origin of the point $\mu(T=0) = 0$ for the 3D $A1$ phase. We show in what follows that at the point $\mu(T=0) = 0$, we probably deal with a quantum phase transition [17, 18].

On the global phase diagram, the BCS domain with the chemical potential $\mu > 0$ occupies the region of negative values of the gas parameter $\lambda_p = V_p p_F^3 < 0$ (or the negative values of the scattering length $a_p$). It also stretches to small positive values of the inverse gas parameter $1/\lambda_p \leq 1$ and is separated from the BEC domain (where $\mu < 0$ and the inverse gas parameter is large and positive, $1/\lambda_p \geq 1$) by the line $\mu(T) = 0$. In the Feshbach resonance regime, the density of "up" spins $n = p_F^3/6\pi^2$ is usually fixed. Deep inside the BCS domain (for small absolute values of the gas parameter $|\lambda_p| \ll 1$), we have the standard BCS-like formula for the critical temperature of the $A1$ phase:

$$T_{Cp} = 0.1\varepsilon_F e^{-\pi/2|\lambda_p|}, \qquad (2)$$

where the prefactor for the 100% polarized $A1$ phase is defined by second-order diagrams of the Gor'kov and Melik–Barchudarov type [19] and is approximately equal to $0.1\varepsilon_F$ [20].[2]

Deep in the BEC domain ($\lambda_p \ll 1$), the well-known Einstein formula is applicable in the leading approximation for Bose condensation of $p$-wave molecules with the density $n/2$ and mass $2m$:

$$T_{Cp} = 3.31\frac{(n/2)^{2/3}}{2m}. \qquad (3)$$

In the unitary limit, $1/\lambda_p = 0$. Hence, $T_{Cp} \approx 0.1\varepsilon_F$ here, and we are still in BCS regime (see [16]). In the rest of the paper, we consider low temperatures $T \ll T_C$, i.e., we work deep in the superfluid parts of BCS and BEC domains of the $A1$ phase.

## 3. QUASIPARTICLE ENERGY AND NODAL POINTS IN THE $A1$ PHASE

For the standard $s$-wave pairing, the quasiparticle spectrum is given by

$$E_p = \sqrt{\left(\frac{p^2}{2m} - \mu\right)^2 + \Delta_0^2}. \qquad (4)$$

It has no zeroes (no nodes) and therefore the topology of the $s$-wave pairing problem is trivial. But for the triplet $A1$ phase, we have

$$E_p = \sqrt{\left(\frac{p^2}{2m} - \mu\right)^2 + \frac{|\boldsymbol{\Delta} \cdot \mathbf{p}|^2}{p_F^2}}, \qquad (5)$$

where $\boldsymbol{\Delta} = \Delta_0(\mathbf{e}_x + i\mathbf{e}_y)$ is the complex order parameter in the $A1$ phase and $\Delta_0$ is the magnitude of the superfluid gap. In fact, $|\boldsymbol{\Delta} \cdot \mathbf{p}|^2 = \Delta_0^2 p^2 \sin^2\theta = \Delta_0^2[\mathbf{p} \times \mathbf{l}]^2$, where $\mathbf{l} = \mathbf{e}_x \times \mathbf{e}_y$ is the unit vector of orbital momentum (see Fig. 3). We note that $p_F$ is fixed by the fixed density $n$. The angle $\theta$ is between the momentum $\mathbf{p}$ and the orbital momentum quantization axis $\mathbf{l} = \mathbf{e}_z$.

For $\mu > 0$ (the BCS domain), there are two nodes in the spectrum for $p^2/2m = \mu$ and $\theta = 0$ or $\pi$. For $\mu < 0$

---

[2] This calculation was done for the nonpolarized $A$ phase in the case where the $s$-wave scattering is totally suppressed. The calculation for the 100%-polarized $A1$ phase yields only a 10% difference from the result in [20] for the prefactor $0.1\varepsilon_F$.

(the BEC domain), there are no nodes. The important point $\mu = 0$ is a boundary between the totally gapped BEC domain and the BCS domain with two nodes of the quasiparticle spectrum corresponding to the south and north poles in Fig. 3. This point for $T = 0$ is often called the topological quantum phase transition point [21, 22].

## 4. LEGGETT EQUATIONS FOR THE $A1$ PHASE

The Leggett equations for the 100%-polarized $A1$ phase in three dimension are the evident generalization of the standard Leggett equations for the $s$-wave BCS–BEC crossover [16, 23]. The first equation is

$$n = \frac{p_F^3}{6\pi^2} = \int_0^{1/r_0} \frac{p^2 dp}{2\pi^2} \int_{-1}^{1} \frac{dx}{2} \left(1 - \frac{\xi_p}{E_p}\right) \frac{1}{2}, \quad (6)$$

where $\xi_p = (p^2/2m - \mu)$,

$$E_p = \sqrt{\xi_p^2 + \frac{\Delta_0^2 p^2}{p_F^2} \sin^2\theta}$$

is the quasiparticle spectrum, and $x = \cos\theta$. This equation defines the chemical potential $\mu$ for a fixed density $n$.

The momentum distribution for the function $1/2(1 - \xi_p/E_p)$ in (6) is depicted in Fig. 4 for different values of $\mu$ corresponding to the BCS and BEC domains.

The second self-consistency equation defines the magnitude of the superfluid gap $\Delta_0$. It is given by

$$-\pi m \operatorname{Re} \frac{1}{f_{l=1}(2\mu)} = \int_{-1}^{1} \frac{dx}{2} \int_0^{1/r_0} p^4 dp \left\{ \frac{1}{E_p} - \frac{1}{\xi_p} \right\}, \quad (7)$$

where

$$\operatorname{Re} \frac{1}{f_{l=1}(2\mu)} = \left(\frac{1}{V_p} + \frac{4m\mu}{\pi r_0}\right)$$

is the real part of the inverse scattering amplitude in the $p$-wave channel for the total energy $E = 2\mu$ of colliding particles. This energy is relevant for the pairing problem, and hence $f_{l=1}(E)$ must be replaced with $f_{l=1}(2\mu)$ in the Legget equations.

Deep in the BCS domain, the solution of the Leggett equations yields

$$\Delta_0 \sim \varepsilon_F e^{-\pi/2|\lambda_p|} \sim T_{Cp}, \quad \mu \approx \varepsilon_F > 0. \quad (8)$$

In three dimensions, the sound velocity is

$$c_s = \left(\frac{n}{m}\frac{d\mu}{dn}\right)^{1/2} = \frac{v_F}{\sqrt{3}}. \quad (9)$$

For $1/|\lambda_p| = 0$, $\Delta_0 \sim \varepsilon_F$, and hence the unitary limit is still inside the BCS domain.

Deep in the BEC domain,

$$\Delta_0 \approx 2\varepsilon_F \sqrt{p_F r_0} \ll \varepsilon_F \quad \text{for} \quad p_F r_0 \ll 1, \quad (10)$$

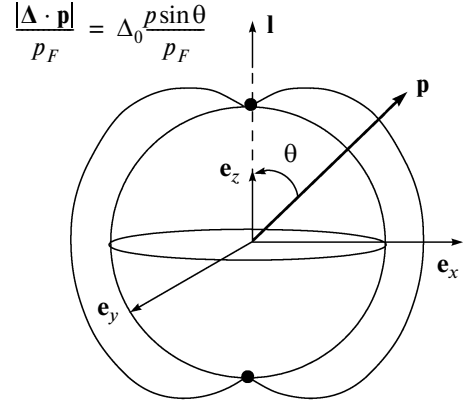

$$\frac{|\Delta \cdot \mathbf{p}|}{p_F} = \Delta_0 \frac{p \sin\theta}{p_F}$$

**Fig. 3.** The topology of the superfluid gap in the $A1$ phase. There are two nodes in the quasiparticle spectrum corresponding to the south and north poles.

and the chemical potential $\mu = -|E_b|/2 + \mu_B/2 < 0$, where, as we already noted,

$$|E_b| = \frac{\pi}{2mr_0 a_p} \quad (11)$$

is the binding energy of a triplet pair (molecule). Accordingly,

$$\mu_B \approx \frac{4\varepsilon_F}{3}\sqrt{p_F r_0} \quad (12)$$

is a bosonic chemical potential that governs the repulsive interaction between two $p$-wave molecules [16]. The sound velocity deep in the BEC domain is given by

$$c_s = \left(\frac{n_B}{2m}\frac{d\mu_B}{dn_B}\right)^{1/2} \approx \frac{v_F}{\sqrt{3}}\sqrt{p_F r_0} \ll v_F \quad (13)$$

$$\text{for} \quad p_F r_0 \ll 1,$$

where $n_B = n/2$ is the bosonic density.

As $\mu \longrightarrow 0$ (more rigorously, for $|\mu| < \Delta_0^2/\varepsilon_F$), we have

$$\Delta_0(\mu = 0) = 2\varepsilon_F\sqrt{p_F r_0} \quad (14)$$

for the magnitude of the superfluid gap.

For the gas parameter $\lambda_p$ at the point $\mu = 0$, we have

$$\lambda_p(\mu = 0) = \frac{3\pi}{4} > 0. \quad (15)$$

Hence, the interesting point $\mu = 0$ is effectively in the BEC domain (in the domain of positive $p$-wave scattering lengths $a_p > 0$). Accordingly, for $\mu = 0$, the binding energy is

$$|E_b| = \frac{4}{3}\varepsilon_F p_F r_0. \quad (16)$$

The sound velocity squared for $\mu = 0$ is given by

$$c_s^2 = \frac{v_F^2}{3}p_F r_0 \quad (17)$$

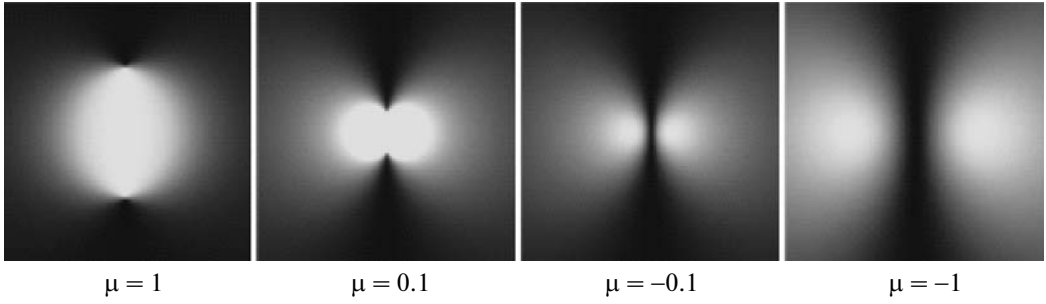

$\mu = 1$  $\mu = 0.1$  $\mu = -0.1$  $\mu = -1$

**Fig. 4.** Schematic momentum distribution of the function $(1 - \xi_p/E_p)/2$ entering (6) in the $(p_x, p_z)$ plane for $p_y = 0$, $\Delta_0 = 1$, and $E_F = p_F^2/2m$ in the BCS–BEC crossover for the 3D $A1$ phase. The different values of $\mu$ correspond to the situation deep in the BCS domain ($\mu = 1$), deep in the BEC domain ($\mu = -1$), and in the important region close to $\mu = 0$ ($\mu = +0.1$ and $\mu = -0.1$).

and coincides with (10), obtained deep in the BEC domain. A careful analysis of Leggett equations close to $\mu = 0$ shows that the derivative $\partial \Delta / \partial \mu$ also has no singularities at this point. The second derivative $\partial^2 n / \partial \mu^2$ is also continuous at $\mu = 0$, and hence the anomaly appears only in higher derivatives, in qualitative agreement with the numerical calculations in [24] in three dimensions.

At the same time, the careful analysis of compressibility in the two-dimensional case [25] shows continuous behavior but with a kink already developed in $\partial n / \partial \mu$ in the 100%-polarized $(p_x + ip_y)$ phase for $\mu = 0$ on the level of analytic as well as numerical calculations [24]. To be more specific,

$$\frac{\partial n}{\partial \mu} \propto 1 + \frac{\mu \varepsilon_F}{\Delta_0^2}[1 - \operatorname{sgn}\mu] \quad (18)$$

and hence $\partial n/\partial \mu \propto 1$ as $\mu \longrightarrow +0$, and $\partial n/\partial \mu \propto 1 + 2\mu\varepsilon_F/\Delta_0^2$ as $\mu \longrightarrow -0$.

## 5. SPECIFIC HEAT AND NORMAL DENSITY AT LOW TEMPERATURES $T \ll T_C$

In this section, we study the thermodynamic functions, the normal density $\rho_n$, and the specific heat $C_v$ in three-dimensional resonant $p$-wave superfluids with the $A1$ symmetry at low temperatures $T \ll T_C$. Our goal is to find nontrivial contributions to $\rho_n$ and $C_v$ from the nodal points on the mean field level.

### 5.1. Specific Heat in the Three-Dimensional A1 Phase

The fermionic (quasiparticle) contribution to $C_v$ at the mean-field level in three dimensions is (see [26])

$$C_v = \int \frac{\partial n_0(E_p/T)}{\partial T} E_p \frac{d^3\mathbf{p}}{(2\pi)^3}, \quad (19)$$

where $n_0(E_p/T) = (e^{E_p/T} + 1)^{-1}$ is the quasiparticle distribution function and $E_p$ is the quasiparticle energy given by (5).

The result of the calculation is

$$C_v \sim N(0) \frac{T^3}{\Delta_0^2} \quad (20)$$

deep in the BCS domain, where $N(0) = mp_F/2\pi^2$ is the density of states at the Fermi surface. Deep in the BEC domain, $C_v$ is given by an exponential,

$$C_v \sim \frac{(2mT)^{3/2}}{2\pi^2} \frac{E_b}{4T^2} e^{-|E_b|/2T}, \quad (21)$$

with $|E_b|$ in (11).

Finally, in the interesting region of small $\mu$ and low temperatures ($|\mu| \ll T \ll \Delta_0^2/\varepsilon_F$ and hence in the classical limit $|\mu|/T \longrightarrow 0$), we have a notrivial temperature dependence for $C_v$:

$$C_v \sim \frac{(2mT)^{3/2}}{2\pi^2} \frac{\varepsilon_F T}{\Delta_0^2}. \quad (22)$$

We note that in the opposite quantum limit $T/|\mu| \longrightarrow 0$ ($T \ll |\mu| \ll \Delta_0^2/\varepsilon_F$), we have

$$C_v \sim \frac{1}{2\pi^2} \frac{\varepsilon_F T}{\Delta_0^2} \frac{T^2 m^{3/2}}{\mu^{1/2}} \quad (23)$$

in the BCS domain and

$$C_v \sim \frac{(2mT)^{3/2}}{2\pi^2} \frac{|\mu|^3}{T^3} e^{-|\mu|/T} \frac{\varepsilon_F T}{\Delta_0^2} \quad (24)$$

in the BEC domain. In this limit, $C_v$ behaves very differently in the BCS and BEC domains.

For $|\mu| \sim T$, results (23) and (24) coincide with (22) by the order of magnitude.

For small $|\mu|$, but intermediate temperatures $|\mu| \ll \Delta_0^2/\varepsilon_F \ll T \ll \Delta_0$, we recover a more expected result:

$$C_v \sim \frac{(2mT)^{3/2}}{2\pi^2}. \qquad (25)$$

But the bosonic contribution (the contribution from sound waves) prevails at these temperatures and yields

$$C_v^B \sim \frac{T^3}{c_s^3}\frac{1}{2\pi^2}, \qquad (26)$$

where the sound velocity $c_s$ is given by (9) in the BCS domain, and by (13) and (17) in the BEC domain and close to $\mu = 0$.

We see that a power-law fermionic contribution $C_v \propto T^{5/2}$ at low temperatures and $C_v \propto T^{3/2}$ at intermediate temperatures can be separated from the bosonic contribution $C_v^B \propto T^3$ close to the important point $\mu = 0$. We also see very different behaviors of $C_v$ in the BCS and BEC domains in the limit $T/|\mu| \to 0$.

Analogously, in the two dimensional 100%-polarized $(p_x + ip_y)$-phase in the quantum limit $T \ll |\mu| \ll \Delta_0^2/\varepsilon_F (T/|\mu| \to 0)$, the quasiparticle contribution is given by

$$C_v \sim \frac{1}{2\pi}\frac{m\varepsilon_F}{\Delta_0^2}T^2 \qquad (27)$$

in the BCS domain for $\mu \to +0$. We note that the phonon contribution has the same order of magnitude as the fermionic contribution in the BCS domain. In the BEC domain for $\mu \to -0$,

$$C_v \sim \frac{1}{2\pi}\frac{m\varepsilon_F|\mu|^3}{\Delta_0^2 T}e^{-|\mu|/T}. \qquad (28)$$

We note that in both three and two dimensions for $T \neq 0$, we are effectively always in the classical limit $|\mu|/T \to 0$, because $\mu$ is continuous close to $\mu = 0$. Hence, the real phase transition occurs only at $T = 0$ [21, 22].

### 5.2. Normal Density in the Three-Dimensional A1 Phase

The quasiparticle contribution to the normal density in the 3D A1 phase is (see [26])

$$\rho_n = -\frac{1}{3}\int p^2\frac{\partial n_0(E_p/T)}{\partial E_p}\frac{d^3\mathbf{p}}{(2\pi)^3}. \qquad (29)$$

Deep in the BCS domain, the evaluation of $\rho_n$ yields

$$\rho_n \sim \rho\frac{T^2}{\Delta_0^2}, \qquad (30)$$

where $\rho = mn$ is the total mass density. We note that rigorously speaking, Eq. (30) yields the longitudinal component of the normal density tensor $\rho_{nl}$. There is also a small transverse contribution $\rho_{nt} \sim T^4$ first obtained in [21].

Deep in the BEC domain, the normal density is exponential,

$$\rho_n \sim \frac{m}{\pi^2}(2mT)^{3/2}e^{-|E_b|/2T}. \qquad (31)$$

Finally, close to $\mu = 0$ at low temperatures ($|\mu| \ll T \ll \Delta_0^2/\varepsilon_F$ and hence in the classical limit $|\mu|/T \to 0$), we have

$$\rho_n \sim \frac{m}{\pi^2}(2mT)^{3/2}\frac{\varepsilon_F T}{\Delta_0^2}. \qquad (32)$$

In the opposite quantum limit $T/|\mu| \to 0$ ($T < |\mu| < \Delta_0^2/\varepsilon_F$), we have

$$\rho_n \sim \frac{m}{\pi^2}\frac{\varepsilon_F T}{\Delta_0^2}2mT(2m|\mu|)^{1/2} \qquad (33)$$

in the BCS domain and

$$\rho_n \sim \frac{m}{\pi^2}\frac{\varepsilon_F T}{\Delta_0^2}e^{-|\mu|/T}2m|\mu|(2mT)^{1/2} \qquad (34)$$

in the BEC domain, and therefore the behavior of $\rho_n$ is again very different in the BCS and BEC domains in the quantum limit.

For $|\mu| \sim T$, results (33) and (34) coincide with (32) by the order of magnitude.

At intermediate temperatures $|\mu| \ll \Delta_0^2/\varepsilon_F \ll T \ll \Delta_0$, the normal density yields

$$\rho_n \sim \frac{m}{\pi^2}(2mT)^{3/2}, \qquad (35)$$

as expected. But the bosonic (phonon) contribution from the sound waves prevails at these temperatures and yields (see [26])

$$\rho_n^B \sim \frac{T^4}{c_s^5}, \qquad (36)$$

where $c_s$ is again respectively given by (5), (13), and (17) in the BCS and BEC domains and close to $\mu = 0$. We can again separate the fermionic (quasiparticle) contribution to $\rho_n$ ($\rho_n \propto T^{5/2}$ at low temperatures and $\rho_n \propto T^{3/2}$ at intermediate temperatures) from the bosonic contribution ($\rho_n \propto T^4$) close to $\mu = 0$. We also see very different behaviors of $\rho_n$ in the BEC and BCS domains in the quantum limit $T/|\mu| \to 0$. The same behavior holds in the two dimensional case.

## 6. ORBITAL WAVES, INTRINSIC ANGULAR MOMENTUM AND CHIRAL ANOMALY IN THE $A1$ PHASE

Topological effects in the $A1$ phase are already pronounced in the spectrum of orbital waves and in the superfluid hydrodynamics at low temperatures $T \rightarrow 0$, especially in the BCS domain. There, by symmetry requirements, we can write the total mass current as

$$\mathbf{j}_{\text{tot}} = \mathbf{j}_B + \mathbf{j}_{\text{an}}, \quad (37)$$

where

$$\mathbf{j}_{\text{an}} = -\frac{\hbar}{4m} C_0 (\mathbf{l} \cdot \text{curl}\, \mathbf{l}) \mathbf{l} \quad (38)$$

is an anomalous current. In the BEC domain, $C_0 = 0$ and the anomalous current is absent. This is because

$$\frac{N(0)}{2} \int d\xi_p \left(1 - \frac{\xi_p}{|\xi_p|}\right) = 0$$

in the BEC domain (for $\xi_p > 0$), while this integral is nonzero and defines the total density in the BCS domain. However, it is a difficult question whether $C_0 = 0$ in the BCS domain.

At the same time, $\mathbf{j}_B$ in (37) is the total mass current in the BEC domain for $p$-wave molecules. It is given by

$$\mathbf{j}_B = \rho \mathbf{v}_s + \frac{\hbar}{2m} \text{curl}\, \frac{\rho \mathbf{l}}{2}, \quad (39)$$

where $\mathbf{L} = \hbar \rho \mathbf{l}/2m$ is the density of orbital momentum and $\mathbf{v}_s$ is the superfluid velocity.

The anomalous current $\mathbf{j}_{\text{an}}$ violates the conservation law for the total mass current (total linear momentum) $\mathbf{j}_{\text{tot}}$ hot because it cannot be expressed as a divergence of a momentum tensor $\Pi_{ik}$:

$$\frac{\partial j^i_{\text{tot}}}{\partial t} \neq -\frac{\partial}{\partial x_k}(\Pi_{ik}). \quad (40)$$

Therefore, the presence of an anomalous current destroys the superfluid hydrodynamics of the $A1$ phase as $T \rightarrow 0$. Its contribution to the equation for the total linear momentum (to $\partial j^i_{\text{tot}}/\partial t$) can be compensated only by adding a term with the relative normal velocity and normal density $\rho_n(T=0)(\mathbf{v_n} - \mathbf{v_s})$ to the total current $\mathbf{j}_{\text{tot}}$ already at $T = 0$ (see [5, 6]). The anomalous current also significantly changes the spectrum of orbital waves. This additional Goldstone branch of collective excitations in the $A1$ phase is related to the rotation of the $\mathbf{l}$ vector around a perpendicular axis. It is quadratic at low frequencies (the $A1$ phase is called an orbital ferromagnet; it is also a spin ferromagnet). However, the coefficient at $q^2$ is drastically different in the BCS and BEC domains.

In the BEC domain for small $\omega$ and $\mathbf{q}$, $\rho \omega \sim \rho q_z^2/m$ or, equivalently,

$$\omega \sim q_z^2/m. \quad (41)$$

But in the BCS domain,

$$(\rho - C_0)\omega \sim \rho \frac{q_z^2}{m} \ln \frac{\Delta_0}{v_F |q_z|}. \quad (42)$$

The most straightforward way to obtain (42) is to use the diagram technique in [27] for the collective excitation spectrum in $p$-wave and $d$-wave superfluids. The solution of the Bethe–Salpeter integral equation for the Goldstone spectrum of orbital waves in the approach of [27] involves the Ward identity between the total vertex $\Gamma$ and the self-energy $\Sigma$, which is based on the generator of rotations of the $\mathbf{l}$ vector around perpendicular axis. In the general form, for small $\omega$ and $\mathbf{q} = q_z \mathbf{e}_z$ it is given by

$$\int_{-1}^{1} \frac{d\cos\theta}{2} \cos^2\theta$$
$$\times \int \frac{p^2 dp}{2\pi^2} \left[\frac{\omega^2}{8E_p^3} + \frac{\omega \xi_p}{4E_p^3} - \frac{p_z^2}{m^2} q_z^2 \frac{1}{4E_p^3}\right] = 0. \quad (43)$$

Deep in the BCS domain (for $\mu \approx \varepsilon_F > 0$), we can replace $p^2 dp/2\pi^2$ with $N(0) d\xi_p$ (where $N(0) = mp_F/2\pi^2$) and $p_z^2/m^2$ with $v_F^2 \cos^2\theta$. This yields

$$N(0) \int_{-1}^{1} \frac{d\cos\theta}{2} \cos^2\theta \int d\xi_p \left[\frac{\omega^2}{8E_p^3} + \frac{\omega \xi_p}{4E_p^3} - \frac{v_F^2 \cos^2\theta q_z^2}{4E_p^3}\right]. \quad (44)$$

Using the estimates

$$\int_{-\varepsilon_F}^{\infty} \frac{d\xi_p}{E_p^3} = \frac{1}{\Delta_0^2 \sin^2\theta} \quad (45)$$

and

$$\int_{-\varepsilon_F}^{\infty} \frac{\xi_p d\xi_p}{E_p^3} \approx \frac{1}{\varepsilon_F}, \quad (46)$$

we obtain

$$N(0) \left\{ \frac{\omega^2}{\Delta_0^2} \ln \frac{\Delta_0}{\omega} + \frac{\omega}{\varepsilon_F} - \frac{v_F^2 q_z^2}{\Delta_0^2} \ln \frac{\Delta_0}{v_F |q_z|} \right\} = 0. \quad (47)$$

More rigorously, the equation for the spectrum is biquadratic due to rotation of the $\mathbf{l}$ vector, as it should be for bosonic excitations:

$$\left(\frac{\omega^2}{\Delta_0^2} \ln \frac{\Delta_0}{\omega} + \frac{\omega}{\varepsilon_F}\right)^2 \sim \left(\frac{v_F^2 q_z^2}{\Delta_0^2} \ln \frac{\Delta_0}{v_F |q_z|}\right)^2. \quad (48)$$

For small frequencies $\omega < \Delta_0^2/\varepsilon_F$, the spectrum is quadratic:

$$\omega \frac{\Delta_0^2}{\varepsilon_F} = v_F^2 q_z^2 \ln \frac{\Delta_0}{v_F |q_z|} \quad (49)$$

or, equivalently,

$$\omega \frac{\Delta_0^2}{\varepsilon_F^2} = \frac{q_z^2}{m} \ln \frac{\Delta_0}{v_F |q_z|}. \quad (50)$$

Hence, comparing (50) and (42), we obtain

$$\frac{\rho - C_0}{\rho} = \frac{\Delta_0^2}{\varepsilon_F^2} \ll 1, \quad (51)$$

and therefore $C_0 \approx \rho$ deep in the BCS domain.

In superfluid $^3$He-$A$, for example, $\Delta_0/\varepsilon_F \sim T_c/\varepsilon_F \sim 10^{-3}$ [12], and hence $(\rho - C_0)/\rho \sim 10^{-6}$.

At the same time, for larger frequencies $\Delta_0^2/\varepsilon_F < \omega < \Delta_0$, the spectrum is almost linear:

$$\omega^2 \ln \frac{\Delta_0}{\omega} = v_F^2 q_z^2 \ln \frac{\Delta_0}{v_F |q_z|}. \quad (52)$$

Deep in the BEC domain for $\mu \approx -|E_b|/2 < 0$, it follows from (43) that

$$\omega^2 + |\mu|\omega \sim |\mu|\frac{q_z^2}{m}. \quad (53)$$

Of course, the exact equation is again biquadratic due to rotation,

$$(\omega^2 + |\mu|\omega)^2 \sim (|\mu|q_z^2/m)^2. \quad (54)$$

Hence, $\omega \sim q_z^2/m$ for $\omega < |\mu|$ in agreement with (41). Moreover, this means that $(\rho - C_0)/\rho = 1$ deep in the BEC domain, and therefore $C_0 = 0$.

The same estimate for the density of the intrinsic angular momentum yields

$$\mathbf{L} = (\rho - C_0)\mathbf{l}\frac{\hbar}{2m}$$

for the BCS domain and $\mathbf{L} = \rho\mathbf{l}\hbar/2m$ for the BEC domain. We note that even in the BCS case, different calculations yield different results. For $\mathbf{l}$ = const, the evaluation in [28, 29] yields $\mathbf{L} = \rho\mathbf{l}\hbar/2m$, while the inclusion of inhomogeneous textures of the $\mathbf{l}$ vector restores the expression

$$\mathbf{L} = (\rho - C_0)\mathbf{l}\frac{\hbar}{2m}.$$

We note that according to Leggett [30], the total $N$-particle Hamiltonian $\hat{H}$ exactly commutes with the $z$-projection of the angular momentum $\hat{L}_z = \hbar \hat{N}/2$. This fact is in favor of the result $\mathbf{L} = \rho\mathbf{l}\hbar/2m$ for $\mathbf{l}$ = const in the BCS domain. Returning to the complicated problem of the chiral anomaly, we reconsider the two different approaches to this problem worked out in the late 1980s.

## 7. TWO DIFFERENT APPROACHES TO THE CHIRAL ANOMALY PROBLEM IN THE $A1$ PHASE

The first approach [4] is based on supersymmetric hydrodynamics of the $A1$ phase.

### 7.1. Supersymmetric Hydrodynamics of the A1 Phase

The idea in [4] was to check whether the chiral anomaly (more precisely, the term $j_{an}\mathbf{v}_s$ in the total energy) is directly related to the zeroes of the gap. The authors of [4] assumed that in a condensed matter system at low frequencies, the only physical reason for an anomaly can be related to the infrared singularity. We note that ultraviolet singularities are absent in condensed matter systems, in contrast to quantum electrodynamics. Strong (critical) fluctuations are also suppressed in three-dimensional systems. The main idea in [4] was therefore to check the dangerous infrared regions where the gap is practically zero. For that, the authors of [4] considered the total hydrodynamic action $S_{tot}$ of the $A1$ phase for low frequencies and small $q$ vectors as a sum of bosonic and fermionic contributions,

$$S_{tot} = S_B + S_F, \quad (55)$$

where $S_B(\rho, \mathbf{l}, v_s)$ is the bosonic action and $S_F$ is the fermionic action related to the zeroes of the superfluid gap (see Fig. 5).

Generally speaking, the idea in [4] was to use super-symmetric hydrodynamics to describe all the zero-energy Goldstone modes, including the fermionic Goldstone mode that comes from the zeroes of the gap.

The authors of [4] were motivated by the nice paper [31], where the massless fermionic neutrino was for the first time included in the effective infrared Lagrangian for electroweak interactions.

After the integration over fermionic variables, the authors of [4] obtained the effective bosonic action and checked what infrared anomalies were present in it. As a result, they obtained

$$S_B^{eff} = S_B + \Delta S_B, \quad (56)$$

where the nodal contribution to the liquid-crystal-like part of the effective action [32], which is related to the gradient orbital energy, is

$$\Delta S_B = -\frac{p_F^2 v_l}{32\pi^2} \int d^4x \\ \times \left\{ [\mathbf{l} \times \text{curl}\mathbf{l}]^2 + \frac{v_l^2}{v_l^2}(\mathbf{l} \cdot \text{curl}\mathbf{l})^2 \right\} \left( \ln \frac{l_{MF}^2}{r^2} \right). \quad (57)$$

Here, $x = (\mathbf{r}, t)$, $l_{MF}$ is the mean free path, and $\xi_0 < r < l_{MF}$ ($\xi_0 \sim v_F/\Delta_0$ is the coherence length).

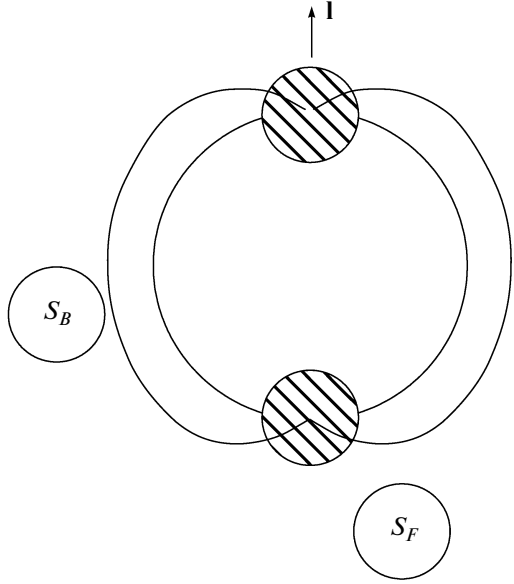

**Fig. 5.** A qualitative illustration of the fermionic ($S_F$) and bosonic ($S_B$) contributions to the total hydrodynamic action $S_{\text{tot}}$ of the $A1$ phase at $T \longrightarrow 0$.

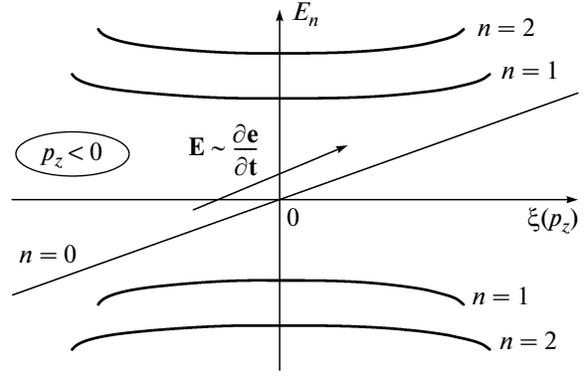

**Fig. 6.** Level structure of the Dirac equation in the magnetic field $B = \mathbf{l} \cdot \text{curl}\mathbf{l}$. All the levels with $n \neq 0$ are doubly degenerate. The zeroth level is chiral. It crosses the origin at $|p_z| = p_F$ in the BCS domain ($\mu > 0$). We also illustrate the concept of the spectral flow, which is to be discussed in Section 9.

Expression (57) for $\Delta S_B$ has a general character and is valid in both weak-coupling and strong-coupling limits.

We note that $v_t \sim v_F \Delta_0/\varepsilon_F \ll v_F$, and $v_l \sim v_F$ in the weak-coupling case. It follows that only weak logarithmic singularities are present in $\Delta S_B$.

However, we do not observe any sign of a strong singularity (which should actually be δ-functional because the fermionic density $\rho_F$ coming from the nodal regions in $S_F$ is small in comparison with the total density $\rho$). In the other words, we do not see any trace of the anomalous contribution

$$\mathbf{j}_{\text{an}} \cdot \mathbf{v}_s = -\frac{\hbar}{4m} C_0 (\mathbf{l} \cdot \text{curl}\mathbf{l})(\mathbf{l} \cdot \mathbf{v}_s) \qquad (58)$$

in $\Delta S_B$.

Hence, even if the chiral anomaly exists in the BCS domain of the $A1$ phase, it is not directly connected with the dangerous regions of momentum space near zeroes of the gap (it does not have an infrared character).

## 8. THE DIFFERENT APPROACH BASED ON A FORMAL ANALOGY WITH QUANTUM ELECTRODYNAMICS

The authors of [5, 6] proposed a different, and also rather nice approach based on a formal analogy between the anomalous current in $^3$He-$A$ and the chiral anomaly in QED. They assume that the anomalous current with the coefficient $C_0 \sim \rho$ in the BCS domain of the $A1$ phase is not directly related to the zeroes of the gap (and hence is not contained even in the super-symmetric hydrodynamics). They believe that it is related to global topological considerations, and therefore a topological term must be added to the supersymmetric hydrodynamics. To illustrate this point, they solve the microscopic Bogoliubov–de Gennes (BdG) equations for fermionic quasiparticles in a given twisted texture ($\mathbf{l} \parallel \text{curl}\mathbf{l}$) of the $\mathbf{l}$ vector. To be more specific, they consider the case

$$\mathbf{l} = \mathbf{l}_0 + \delta\mathbf{l} \qquad (59)$$

with

$$l_z = l_{0z} = e_z, \quad l_y = \delta l_y = Bx, \quad l_x = 0, \qquad (60)$$

where $\mathbf{e}_z$ is the direction of a nonperturbed $\mathbf{l}$ vector. In this case,

$$\mathbf{l} \cdot \text{curl}\mathbf{l} = l_x \frac{\partial l_y}{\partial x} = B = \text{const} \qquad (61)$$

and, accordingly,

$$\mathbf{j}_{\text{an}} = -\frac{\hbar}{4m} C_0 B \mathbf{e}_z. \qquad (62)$$

After linearization, the BdG equations become equivalent to the Dirac equation in the homogeneous magnetic field $B = (\mathbf{l} \cdot \text{curl}\mathbf{l})$. Solving the Dirac equation yields the level structure for fermionic quasiparticles

$$E_n(p_z) = \pm\sqrt{\xi^2(p_z) + \tilde{\Delta}_n^2}, \qquad (63)$$

where $\xi(p_z) = p_z^2/2m - \mu$, $e = p_z/p_F = \pm 1$ is the electric charge, and

$$\tilde{\Delta}_n^2 = 2n v_t^2 p_F |eB| \qquad (64)$$

is the gap squared, with $v_t \sim v_F \Delta_0/\varepsilon_F$.

For $n \neq 0$ (see Fig. 6), all the levels are gapped, $\tilde{\Delta}_n \neq 0$, and are doubly degenerate with respect to $p_z \longrightarrow -p_z$. Their contribution to the total mass current is zero as $T \longrightarrow 0$.

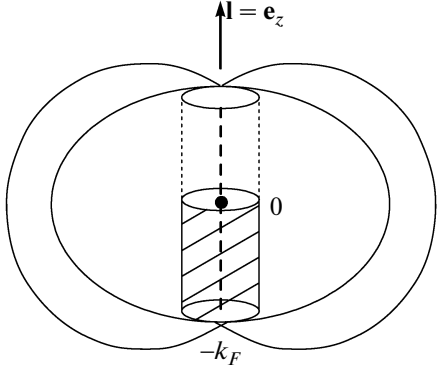

**Fig. 7.** The contribution to the coefficient $C_0$ is governed by a narrow cylindrical tube of the length $p_F$ and width $\langle p_y^2 \rangle \sim p_F |eB|$ inside the Fermi sphere.

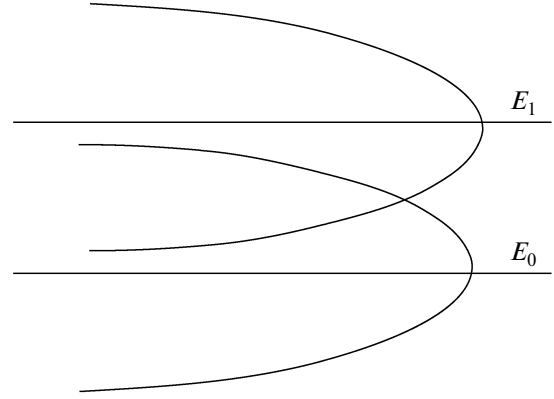

**Fig. 8.** The possible role of damping in reaching the hydrodynamic limit for low frequencies and small **k** vectors for $\gamma > \omega_0$ ($\omega_0 = E_1 - E_0$ is the level spacing).

For $n = 0$, there is no gap ($\tilde{\Delta}_0 = 0$), and we have an asymmetric chiral branch that exists only for $p_z < 0$ or, more precisely, for one sign of $eB$ (see [5] for more details). The energy spectrum for $n = 0$ is given by

$$E_0 = \xi(p_z). \qquad (65)$$

We can say that there is no gap for the zeroth Landau level. Moreover, in the BCS domain, $E_0 = 0$ for $|p_z| = p_F$, which means that the chiral level crosses the origin in Fig. 6 and we have a zero mode.

We note that in the BEC domain, $E_0 \geq |\mu|$ and the zeroth Landau level does not cross the origin. The absence of a zero mode in the BEC domain is the physical reason why $C_0 = 0$ there.

The zeroth Landau level gives an anomalous contribution to the total current in the BCS domain:

$$\mathbf{j}_{an}(\mathbf{r} = 0) = -\mathbf{e}_z (\mathbf{l} \cdot \text{curl}\mathbf{l}) \int_{p_z < 0} \frac{p_z}{2\pi^2} d\xi(p_z) \qquad (66)$$
$$= -\frac{\hbar C_0}{4m} (\mathbf{l} \cdot \text{curl}\mathbf{l})\mathbf{l},$$

where

$$\frac{(\mathbf{l} \cdot \text{curl}\mathbf{l})p_z}{2\pi^2 p_F} = \frac{eB}{2\pi^2} = \int |f_0|^2 \frac{dp_y}{2\pi}, \qquad (67)$$

and hence

$$C_0 \approx \frac{mp_F^3}{6\pi^2} \approx \rho \qquad (68)$$

in the BCS domain.

We note that $f_0(x - p_y/eB)$ in (67) is an eigen-function for the zeroth Landau level. It is easy to see that the integral for $C_0$ in (66) and (67) is governed by the narrow cylindrical tube inside the Fermi sphere (see Fig. 7) with the length $p_F$ parallel to the **l** vector and with the radius of the cylinder squared given by

$$\langle p_y^2 \rangle \sim p_F |eB|. \qquad (69)$$

According to the ideas in [5, 18], this tube plays the role of a vortex in momentum space, thus providing a normal core and anomalous current at $T = 0$.

We note that a key result in [5, 6] related to the absence of the gap for the energy of the zeroth Landau level (see Eq. (65)) is quite stable with respect to small modifications of the texture of the **l** vector in Eq. (60). Our careful analysis shows that the account of small bending corrections with $[\mathbf{l} \cdot \text{curl}\mathbf{l}] \neq 0$ (small tilting of the magnetic field with respect to the $(x, y)$ plane $\mathbf{B} = B_0 \mathbf{e}_z + B_1 \mathbf{e}_x$) as well as of small inhomogeneties of a magnetic field $B = B_0 + B_1 x$, which lead to a double-well effective potential, does not suppress the zero mode in the spectrum of the BdG equation (does not lead to the appearance of a gap $\tilde{\Delta}_{n=0}$ for the zeroth Landau level).

## 9. HOW TO REACH THE HYDRODYNAMIC REGIME $\omega\tau \ll 1$

In spite of the zero mode stability, the authors of [4] expressed their doubts regarding the calculation of $C_0$ based on the Dirac equation in the homogeneous magnetic field $B = \mathbf{l} \cdot \text{curl}\mathbf{l}$. From their standpoint, the calculation of $C_0$ from (66) and (67) is an oversimplification of a complicated many-particle problem. In particular, they emphasized the role of the finite damping $\gamma = 1/\tau$ and of the other residual interactions in destroying the chiral anomaly, which is connected with the states inside the Fermi sphere, thus restoring the superfluid hydrodynamics (without the normal velocity $\mathbf{v}_n$ and the normal density $\rho_n$). Indeed, if the damping $\gamma$ is larger than the level spacing of the Dirac equation, we have

$$\omega_0 = v_t p_F \sqrt{\frac{|\mathbf{l} \cdot \text{curl}\,\mathbf{l}|}{p_F}} \qquad (70)$$

in the case where $\xi(p_z) = 0$, and then the contribution from the zeroth Landau level should be washed out by the damping (see Fig. 8) and the hydrodynamic regime is established. The damping $\gamma$ for the chiral fermions (for fermions living close to the nodes), in a very clean $A1$ phase without impurities, it is defined at $T = 0$ by the different decay processes (see [26]).

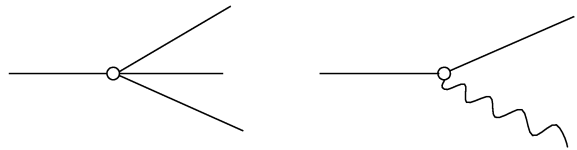

**Fig. 9.** Different decay processes for damping of chiral fermions at $T = 0$: the standard three-fermion decay process and a decay process with an orbital wave emission.

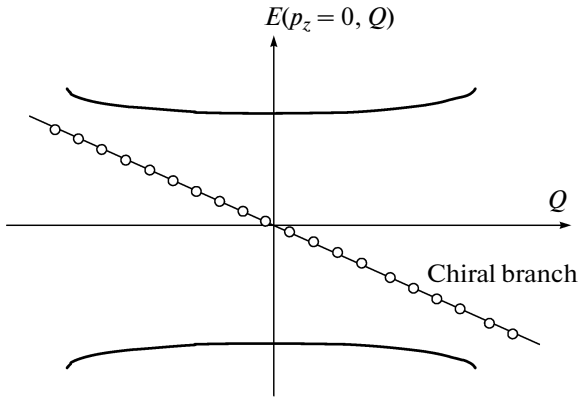

**Fig. 10.** The level structure in the vortex core of $^3$He-$A$. All the branches are even in the generalized angular momentum $Q$, but one branch $E(p_z = 0, Q) = -\omega_0 Q$, which crosses zero energy at $Q = 0$, is chiral (odd in $Q$). It participates in the momentum exchange between the fermions in the vortex core and the heat bath fermions in the hydrodynamic limit $\omega\tau \ll 1$ according to [21].

It is natural to assume that the only parameter that determines $\gamma$ at $T = 0$ for chiral fermions is $\Delta_0 \langle \theta \rangle = \Delta_0 \langle p_\perp \rangle / p_F$. The leading term in decay processes is given by the emission of an orbital wave (see Fig. 9). It is given by

$$\gamma \propto \left[ \frac{\Delta_0^2 p_\perp^2 / p_F^2 + v_F^2 (p_z - p_F)^2}{\varepsilon_F} \right]. \qquad (71)$$

For $p_z = p_F$ ($\xi(p_z) = 0$), we have

$$\gamma \sim \frac{\Delta_0^2 p_\perp^2}{\varepsilon_F p_F^2}. \qquad (72)$$

We note that for the chiral fermions on the zeroth Landau level, we have

$$\frac{\langle p_\perp \rangle}{p_F} = \left( \frac{|\mathbf{l} \cdot \text{curl}\,\mathbf{l}|}{p_F} \right)^{1/2} \qquad (73)$$

and the level spacing for $\xi(p_z) = 0$ is

$$\omega_0 \sim \Delta_0 \frac{\langle p_\perp \rangle}{p_F}. \qquad (74)$$

Hence, $\gamma/\omega_0 \ll 1$ close to the zero mode for these two decay processes, and a ballistic regime is established. It is therefore difficult to wash out the contribution form the zeroth Landau level by the different decay processes in the superclean $^3$He-$A1$ phase at $T = 0$. We note that the hydrodynamic regime $\omega\tau \ll 1$ could be easily reached in the presence of nonmagnetic impurities or in the presence of aerogel [33–35].

### 9.1. The Concept of the Spectral Flow and the Exact Anomaly Cancelation

If the anomalous current exists in a superclean $A1$ phase at $T = 0$, it should be compensated somehow. According to [5], the deficit in the equation for the conservation of the total linear momentum due to the presence of an anomalous current,

$$\frac{\partial j^i_{\text{an}}}{\partial t} + \frac{\partial \Pi_{ik}}{\partial x_k} = I, \qquad (75)$$

where

$$I = \frac{3\hbar}{4m} C_0 \mathbf{l} \cdot \left( \text{curl}\,\mathbf{l} \times \frac{\partial \mathbf{l}}{\partial t} \right),$$

is exactly compensated by the quasiparticle contribution $\mathbf{P}_{\text{quas}}$:

$$\frac{\partial \mathbf{P}^i_{\text{quas}}}{\partial t} + \frac{\partial \Phi_{ik}}{\partial x_k} = -I, \qquad (76)$$

where $\mathbf{P}_{\text{quas}} = \rho_n(T = 0)(\mathbf{v}_n - \mathbf{v}_s)$ in the hydrodynamic regime.

We note that $\rho_n(T=0) \sim |\mathbf{l} \times \text{curl}\,\mathbf{l}|/\Delta_0$ is a non-analytic function and is related to the nonzero bending. The arguments in [5] are connected with the nonconservation of the axial current $j_5$ in QED, which just compensates $I$ via the Schwinger term $\mathbf{E} \cdot \mathbf{B} \sim \partial \mathbf{l}/\partial t \cdot \text{curl}\,\mathbf{l}$. Physically, according to [5, 36], this cancelation is due to the spectral flow from the negative to the positive energy values along the anomalous branch with $n_L = 0$ in Fig. 6 and then to the quasiparticle bath in the presence of an electric field $\mathbf{E} \sim \partial \mathbf{l}/\partial t$ (of a time-dependent texture of the $\mathbf{l}$ vector).

We note that there is one anomalous level that crosses the zero energy in the physics of a vortex core in the case of cylindrical symmetry (see Fig. 10). At $T = 0$, as a function of the generalized angular momentum $Q$, it represents the set of discrete points separated by a minigap $\omega_0 \sim \Delta_0^2/\varepsilon_F$. Therefore, at $T = 0$ and in the superclean case $\gamma = 1/\tau \longrightarrow 0$, the spectral flow from negative to positive energies is totally suppressed. Thus the Thouless result [37] for the Berry phase without the anomaly is restored for the physics of the vortex friction. An inclusion of a large number of impurities or a finite temperature leads to the revival of the anomaly in the hydrodynamic regime $\omega_0\tau \ll 1$ in the case of vortices. We could therefore assume that the chiral anomaly and the spectral flow are ineffective at $T = 0$ for both vortices and the bulk $A$ phase of the superfluid $^3$He in the superclean limit. Hence, the question of how the total linear momentum is conserved in this case in an infinite system (without walls) is very nontrivial and unresolved so far.

We think that the exact cancelation between the time derivatives of the anomalous and quasiparticle currents should be demonstrated explicitly by deriving and solving the kinetic equations for the nodal quasiparticles in both the ballistic and hydrodynamic regimes. We note that if $T \neq 0$ (as we always have in real experiments), and for low frequencies $\omega\tau(T \neq 0) \ll 1$ ($\tau(T \neq 0) \sim \alpha T^{-n}$), the relative normal velocity $\mathbf{v}_n - \mathbf{v}_s = \partial\varepsilon_0/\partial\mathbf{P}_{quas}$ becomes an additional hydrodynamic variable and hence the cancelation of the linear momentum deficit is to occur automatically.

Thus, the problem of the exact anomaly compensation exists only for $T = 0$. We note that an approach based on the kinetic equation for quasiparticles at different temperatures and the impurity concentrations in a vortex core of the $s$-wave superconductors and the superfluid $^3$He was worked out in [38] in the case of a singular vortex.

In the case of nonsingular vortex structures in $^3$He-$A$, we also note papers [36], where the authors consider the scattering of quasiparticles on the walls of the container for a finite systems to obtain a finite $\gamma$ at $T = 0$. The importance of the prehistory of the orbital texture with the spectral flow concept was also stressed in these papers.

## 10. CONCLUSIONS

We solved the Leggett equations and constructed the phase diagram of the BCS–BEC crossover at low temperatures $T \ll T_C$ for the 100%-polarized 3D $A$1 phase. From the evaluation of the low-temperature specific heat and the normal density, we see the indications of a quantum phase transition close to the point $\mu(T = 0) = 0$. At the same time, deep in the BCS and BEC domains, the crossover ideas of Leggett, Nozieres, and Schmitt–Rink work quite well. In these regions, the phase diagram for the $p$-wave resembles the $s$-wave case in gross features. We discussed the complicated problem of the chiral anomaly and the mass current nonconservation in the BCS $A$1 phase at $T = 0$. We presented two different approaches to this problem, based on the supersymmetric hydrodynamics and on the formal analogy with the Dirac equation in QED. We evaluated the damping $\gamma = 1/\tau$ due to the different decay processes in the superclean BCS $A$1 phase at $T = 0$, and found that $\gamma$ is small in comparison with the level spacing $\omega_0$ of the BdG equation. To reach the hydrodynamic regime $\omega\tau \ll 1$, we need a sufficient amount of aerogel or nonmagnetic impurities at $T = 0$. We assumed that in both the hydrodynamic and ballistic regimes at $T = 0$, we have to derive a reliable kinetic equation to explicitly demonstrate the exact cancelation between time derivatives of the anomalous current $\mathbf{j}_{an} = -\hbar/4mC_0\mathbf{l}\cdot\text{curl}\mathbf{l})$ and of the quasiparticle contribution $\mathbf{P}_{quas}$ in the conservation equation for the total linear momentum $\mathbf{j}_{tot}$. We note that for the full theoretical analysis of the problem, other residual interactions different from damping are also important for the nodal fermions. To check whether a chiral anomaly has an infrared manifestation (which was not captured in the approach based on the supersymmetric hydrodynamics in [4]), it will be useful to derive a complete set of Ward identities between the self-energies of chiral fermions $\Sigma$ and the corresponding vertices $\Gamma$. The idea in this approach is to find either a strong infrared singularity or a powerful reexpansion of the quasiparticle spectrum as $\omega, \mathbf{k} \longrightarrow 0$.

We note that the importance of the residual Fermi-liquid-like interactions for the analysis of a half-integer vortex in the three-dimensional $A$ phase of $^3$He was recently emphasized in [39].

We invite the experimentalists to enter this very interesting problem. It will be nice to measure the spectrum and damping of orbital waves in the superfluid $A$ phase of $^3$He at the low temperatures $T \ll T_C$. As we have already discussed, the spectrum is quadratic for low frequencies $\omega < \Delta_0^2/\varepsilon_F$, and contains the intrinsic angular momentum density as a coefficient of the term linear in frequency (see (50) and (51)).

The damping of orbital waves provides an evaluation of the orbital viscosity in $^3$He-$A$ at low temperatures $T \ll T_C$. We note that even in this case, it is an interesting possibility to derive the overdamped (diffusive) character of the spectrum. This possibility is supported theoretically in [40], where the author obtained several overdamped modes in the partially polarized $A$1 phase via the functional integral technique in the hydrodynamic limit of small $\omega$ and $\mathbf{k}$.

Another possibility of an overdamped diffusive spectrum was considered in [41] in the impurity diagram technique [42, 43] for the hydrodynamic regime $\omega\tau \ll 1$ of spin waves in a frustrated two-dimensional AFM. We note that in the opposite high-frequency regime, the spectrum of spin waves is linear.

Here, it is possible to extend the experiments of the orbital inertia and orbital viscosity for nonsingular vortices in the $A$ phase of $^3$He to the low temperatures $T \ll T_C$. Of course, to have the $A$ phase at low temperatures, we need a strong spin polarization.

We also note that a crossover from the ballistic to the hydrodynamic regime $\omega\tau \ll 1$ could occur due to both the aerogel (the nonmagnetic impurities) or a finite temperature $T \neq 0$, which is always present in a real experiment. In the last case, the damping $\gamma \propto T^n$ is temperature dependent.

Finally, to measure the nontrivial topological effects in two dimensions, we propose to perform experiments with a Josephson current between two thin films or two magnetic taps: one with a two-dimensional axial phase and the topological charge $Q = 1$ [44] and the other with the planar two-dimensional phase with $Q = 0$. We hope that it will be possible to directly measure $\Delta Q = 1$ in this type of experiments.

We note that in the 2D axial phase, the $\mathbf{l}$ vector $\mathbf{l} = [\mathbf{e}_x \times \mathbf{e}_y] = \mathbf{e}_z$ is perpendicular to the plane of 2D films. Hence, the orbital waves, connected, as we discussed, with the rotation of the $\mathbf{l}$ vector around a perpendicular axis, are gapped. The sound wave is the only Goldstone mode in gauge orbital sector. Moreover, $\mathbf{l} \perp \operatorname{curl}\mathbf{l}$ (it is impossible to create a twisted texture in two dimensions). Therefore, the anomalous current $\mathbf{j}_{\mathrm{an}} = -\hbar/4mC_0(\mathbf{l} \cdot \operatorname{curl}\mathbf{l})\mathbf{l} = 0$. Hence, there is no problem with the mass current nonconservation at $T = 0$ [25].

Nontrivial topological effects possibly exist in the spin sector [44].

Here, the anomalous spin current was predicted in the presence of an inhomogeneous magnetic field $\mathbf{H}(\mathbf{r})$ for an $^3$He-$A$ film (the BCS phase)

$$j^{\mathrm{spin}}_{\alpha i} \sim Q\varepsilon_{izk}l_z\partial_k H^\perp_\alpha, \qquad (77)$$

where $\mathbf{H}_\perp \cdot \mathbf{d} = 0$ and $\mathbf{d}$ is the spin vector in the 2D $^3$He film.

Another possibility is to measure the contribution of the massless Majorana fermions for the edge states on the surface of superfluid $^3$He-$B$ and a rough wall (or on the surface of a vibrating wire in the Lancaster experiments) [45].

## ACKNOWLEDGMENTS

The authors acknowledge the interesting and useful discussion with A.S. Alexandrov, A.F. Andreev, P.N. Brusov, Yu.M. Bun'kov, A.V. Chubukov, V.V. Dmitriev, A. Golov, V. Gurarie, I.A. Fomin, W. Halperin, Yu. Kagan, W. Ketterle, K.Yu. Kovalev, D.M. Lee, B.E. Meierovich, V.M. Osadchiev, L.P. Pitaevskii, G. V. Shlyapnikov, M. Stone, V.V. Val'kov, and G.E. Volovik and are grateful to the Russian Foundation for Basic Research (grant no. 08-02-00224) for financial support of this work.

## REFERENCES

1. C. Ticknor, C. A. Regal, D. S. Jin, and J. L. Bohn, Phys. Rev. A: At., Mol., Opt. Phys. **69**, 042 712 (2004); C. A. Regal, C. Ticknor, J. L. Bohn, and D. S. Jin, Phys. Rev. Lett. **90**, 053 201 (2003); C. H. Schunck, M. W. Zwierlein, C. A. Stan, S. M. F. Raupach, W. Ketterle, A. Simoni, E. Tiesinga, C. J. Williams, and P. S. Julienne, Phys. Rev. A: At., Mol., Opt. Phys. **71**, 045 601 (2005).
2. Y. Inada, M. Horikoshi, S. Nakajima, M. Kuwata-Gonokami, M. Ueda, and T. Mukaiyama, arXiv:cond-mat 0803.1405; J. Fuchs, C. Ticknor, P. Dyke, G. Veeravalli, E. Kuhnle, W. Rowlands, P. Hannaford, and C. J. Vale, arXiv:cond-mat/0802.3262.
3. J. P. Gaebler, J. T. Stewart, J. L. Bohn, and D. S. Jin, Phys. Rev. Lett. **98**, 200403 (2007).
4. A. F. Andreev and M. Yu. Kagan, Zh. Éksp. Teor. Fiz. **93** (3), 895 (1987) [Sov. Phys. JETP **66** (3), 594 (1987)].
5. A. V. Balatskii, G. E. Volovik, and V. A. Konyshev, Zh. Éksp. Teor. Fiz. **90** (6), 2038 (1986) [Sov. Phys. JETP **63** (6), 1194 (1986)]; G. E. Volovik, Pis'ma Zh. Eksp. Teor. Fiz. **43** (9), 551 (1986) [JETP Lett. **43** (9), 428 (1986)]; G. E. Volovik and V. P. Mineev, Zh. Éksp. Teor. Fiz. **83** (3), 1025 (1982) [Sov. Phys. JETP **56** (3), 579 (1982)].
6. R. Combescot and T. Dombre, Phys. Rev. B: Condens. Matter **33**, 79 (1986); Phys. Rev. B: Condens. Matter **28**, 5140 (1983).
7. H.-Y. Kee, A. Raghavan, and K. Maki, arXiv:cond-mat/0711.0929.
8. R. Roy and C. Kallin, Phys. Rev. B: Condens. Matter **77**, 174 513 (2008); D. A. Ivanov, Phys. Rev. Lett. **89**, 208 (2001).
9. K. S. Novoselov, A. K. Geim, S. V. Morozov, D. Jiang, M. I. Katsnelson, I. V. Grigorieva, S. V. Dubonos and A. A. Firsov, Nature (London) **438**, 197 (2005).
10. N. Tajima, S. Sugawara, M. Tamura, R. Kato, Y. Nishio, and K. Kajita, Europhys. Lett. **80**, 47002 (2007).
11. T. D. C. Bevan, A. J. Manninen, J. B. Cook, H. Alles, J. R. Hook, and H. E. Hall, J. Low Temp. Phys. **109**, 423 (1997); T. D. C. Bevan, A. J. Manninen, J. B. Cook, J. R. Hook, H. E. Hall, T. Vachaspati, and G. E. Volovik, Nature (London) **386**, 689 (1997).
12. D. Vollhardt and P. Woelfle, *The Superfluid Phases of Helium 3* (Taylor and Francis, London, 1990).
13. *Proceedings of the First Euroconference on Anomalous Complex Superconductors, Crete, Greece, 1999*, Ed. by P. B. Littlewood and G. Varelogiannes (North-Holland, Amsterdam, 1999).
14. L. D. Landau and E. M. Lifshitz, *Course of Theoretical Physics*, Vol. 3: *Quantum Mechanics: Non-Relativistic Theory*, 3rd ed. (Nauka, Moscow, 1989; Pergamon, Oxford, 1991).
15. V. Gurarie and L. Radzihovsky, Ann. Phys. (Weinheim) **322**, 2 (2007); Chi-Ho Cheng and S.-K. Yip, Phys. Rev. Lett. **95**, 070404 (2005); F. R. Klinkhamer and G. E. Volovik, Pis'ma Zh. Éksp. Teor. Fiz. **80** (5), 389 (2004) [JETP Lett. **80** (5), 343 (2004)].
16. R. Combescot, X. Leyronas, and M. Yu. Kagan, Phys. Rev. A: At., Mol., Opt. Phys. **73**, 023 618 (2006); I. V. Brodsky, A. V. Klaptsov, M. Yu. Kagan, R. Combescot, and X. Leyronas, Pis'ma Zh. Éksp. Teor. Fiz. **82** (5), 306 (2005) [JETP Lett. **82** (5), 273 (2005)]; Phys. Rev. A: At., Mol., Opt. Phys. **73**, 032 724 (2006);

R. Combescot, M. Yu. Kagan, and S. Stringary, Phys. Rev. A: At., Mol., Opt. Phys. **74**, 042717 (2006).

17. J. A. Hertz, Phys. Rev. B: Solid State **14**, 1165 (1976).

18. A. J. Millis, Phys. Rev. B: Condens. Matter **48**, 7183 (1993).

19. L. P. Gor'kov and T. K. Melik-Barchudarov, Zh. Éksp. Teor. Fiz. **40**, 1452 (1961) [Sov. Phys. JETP **13**, 1018 (1961)].

20. M. Yu. Kagan and A. V. Chubukov, Pis'ma Zh. Éksp. Teor. Fiz. **47** (10), 525 (1988) [JETP Lett. **47** (10), 614 (1988)]; M. A. Baranov, M. Yu. Kagan, and Yu. Kagan, Pis'ma Zh. Éksp. Teor. Fiz. **64** (4), 273 (1996) [JETP Lett. **64** (4), 301 (1996)]; M. A. Baranov, M. Yu. Kagan, and A. V. Chubukov, Int. J. Mod. Phys. B **6**, 2471 (1992); D. V. Efremov, M. S. Mar'enko, M. A. Baranov, and M. Yu. Kagan, Zh. Éksp. Teor. Fiz. **117** (5), 990 (2000) [JETP **90** (5), 861 (2000)].

21. G. E. Volovik, *The Universe in Helium Droplet* (Oxford University Press, Oxford, 2002); Exotic Properties of Superfluid $^3$He) (World Sci., Singapore, 1992); in *Quantum Analogues: From Phase Transitions to Black Holes and Cosmology*, Ed. by W. G. Unruh and R. Schuetzhold, *Springer Lecture Notes in Physics 718* (Springer, Berlin, 2007), p. 31; arXiv:cond-mat/0601372.

22. N. Read and D. Green, Phys. Rev. B: Condens. Matter **61**, 10 267 (2000).

23. A. J. Leggett, in *Modern Trends in the Theory of Condensed Matter Physics* (Springer, Berlin, 1980), p. 13.

24. M. Iskin and C. A. R. Sa de Melo, Phys. Rev. Lett. **96**, 040402 (2006); S. S. Botelho and C. A. R. Sa de Melo, J. Low Temp. Phys. **140**, 409 (2005).

25. M. Yu. Kagan and S. L. Ogarkov, Laser Phys. **18**, 509 (2008); M. Yu. Kagan and D. V. Efremov, Fiz. Nizk. Temp. (Kharkov) **35** (8–9), 779 (2009) [Low Temp. Phys. **35** (8–9), 610 (2009)]; M. Yu. Kagan and S. L. Ogarkov, J. Phys.: Conf. Ser. **150**, 032 037 (2009).

26. L. D. Landau and E. M. Lifshitz, *Course of Theoretical Physics*, Vol. 9: E. M. Lifshitz and L. P. Pitaevskii, *Statistical Physics: Part 2* (Nauka, Moscow, 1973; Butterworth–Heinemann, Oxford, 1980).

27. V. G. Vaks, V. M. Galitskii, and A. I. Larkin, Zh. Éksp. Teor. Fiz. **42**, 1319 (1962) [Sov. Phys. JETP **15**, 914 (1962)].

28. M. Stone and I. Anduaga, Ann. Phys. (Weinheim) **323**, 2 (2008); M. Stone and F. Gaitan, Ann. Phys. (Weinheim) **178**, 89 (1987); M. Stone, Phys. Rev. B: Condens. Matter **54**, 13222 (1996); M. Stone, Physica B (Amsterdam) **280**, 117 (2000); M. Stone and R. Roy, Phys. Rev. B: Condens. Matter **69**, 184511 (2004); F. Gaitan, Phys. Lett. A **151**, 551 (1990).

29. N. D. Mermin and P. Muzikar, Phys. Rev. B: Condens. Matter **21**, 980 (1980).

30. A. J. Leggett, private communication.

31. D. V. Volkov and V. P. Akulov, Pis'ma Zh. Éksp. Teor. Fiz. **16** (11), 621 (1972) [JETP Lett. **16** (11), 438 (1972)].

32. M. C. Cross, J. Low Temp. Phys. **21**, 525 (1975); M. C. Cross, J. Low Temp. Phys. **26**, 165 (1977).

33. V. V. Dmitriev, D. A. Krasnikhin, N. Mulders, V. V. Zavjalov, and D. E. Zmeev, Pis'ma Zh. Éksp. Teor. Fiz. **86** (9), 681 (2007) [JETP Lett. **86** (9), 594 (2007)].

34. J. Elbs, Yu. M. Bunkov, E. Collin, H. Godfrin, and G. E. Volovik, Phys. Rev. Lett. **100**, 215 304 (2008).

35. I. A. Fomin, Pis'ma Zh. Éksp. Teor. Fiz. **88** (1), 65 (2008) [JETP Lett. **88** (1), 59 (2008)]; I. A. Fomin, Pis'ma Zh. Éksp. Teor. Fiz. **77** (5), 285 (2003) [JETP Lett. **77** (5), 240 (2003)].

36. G. E. Volovik, Zh. Éksp. Teor. Fiz. **102** (6), 1838 (1992) [Sov. Phys. JETP **75** (6), 990 (1992)]; Pis'ma Zh. Éksp. Teor. Fiz. **61** (11), 935 (1995) [JETP Lett. **61** (11), 958 (1995)].

37. D. J. Thouless, Ping Ao, and Qian Niu, Phys. Rev. Lett. **76**, 3758 (1996).

38. N. B. Kopnin and M. M. Salomaa, Phys. Rev. B: Condens. Matter **44**, 9667 (1991); N. B. Kopnin, Physica B (Amsterdam) **210**, 267 (1995); N. B. Kopnin, Physica B (Amsterdam) **280**, 231 (2000); N. B. Kopnin, P. S. Soininen, and M. M. Salomaa, Phys. Rev. B: Condens. Matter **45**, 5491 (1992).

39. V. Vakaryak and A. J. Leggett, arXiv:cond-mat/09062631.

40. P. N. Brusov, M. V. Lomakov, and N. P. Brusova, Fiz. Nizk. Temp. (Kharkov) **21** (1), 111 (1995) [Low Temp. Phys. **21** (1), 85 (1995)]; P. N. Brusov and M. V. Lomakov, J. Low Temp. Phys. **85**, 91 (1991); P. N. Brusov and P. P. Brusov, *Collective Excitations in Unconventional Superconductors and Superfluids* (World Sci., Singapore, 2009).

41. A. V. Chubukov, Phys. Rev. B: Condens. Matter **44**, 12 318 (1991); Phys. Rev. B: Condens. Matter **52**, R3840 (1995).

42. A. A. Abrikosov and L. P. Gorkov, Zh. Éksp. Teor. Fiz. **39**, 1781 (1961) [Sov. Phys. JETP **12**, 1243 (1961)].

43. A. I. Larkin, Zh. Éksp. Teor. Fiz. **58**, 1466 (1970) [Sov. Phys. JETP **31**, 784 (1970)].

44. G. E. Volovik, A. Solov'ev, and V. M. Yakovenko, Pis'ma Zh. Éksp. Teor. Fiz. **49** (1), 55 (1989) [JETP Lett. **49** (1), 65 (1989)].

45. G. E. Volovik, arXiv:cond-mat/09075389; arXiv:cond-mat/090953084.